\documentclass[nohyper]{JHEP3}

\usepackage{amsmath,amssymb}
\usepackage{graphicx}
\usepackage{cite}

\newcommand{\beq}{\begin{equation}}
\newcommand{\eeq}{\end{equation}}
\newcommand{\bea}{\begin{eqnarray}}
\newcommand{\eea}{\end{eqnarray}}

\renewcommand{\thefigure}{\Roman{figure}}
\newcommand{\lsim}
{\raise0.3ex\hbox{$\;<$\kern-0.75em\raise-1.1ex\hbox{$\sim\;$}}}
\newcommand{\gsim}
{\raise0.3ex\hbox{$\;>$\kern-0.75em\raise-1.1ex\hbox{$\sim\;$}}}

\title{Cosmological data analysis of $f(R)$ gravity models}

\author{Z. Giron\'es \footnote{girones@ific.uv.es}, 
A. Marchetti \footnote{alida.marchetti@unimi.it},
O. Mena \footnote{omena@ific.uv.es},
C. Pe\~na-Garay \footnote{carlos.penya@ific.uv.es} and 
N. Rius\footnote{nuria@ific.uv.es} 
\\
Depto.\ de F\'{\i}sica Te\'orica,
IFIC, Universidad de
Valencia-CSIC \\ 
Edificio de Institutos de Paterna, Apt. 22085, 46071 Valencia,
Spain}

\abstract{A class of well-behaved modified gravity models with long enough matter domination epoch and a late-time accelerated expansion is confronted with SNIa, CMB, SDSS, BAO and $H(z)$ galaxy ages data, as well as
current measurements of the linear growth of structure. 
We show that the combination of geometrical probes and 
growth data exploited here allows to rule out $f(R)$ gravity models, 
in particular, the logarithmic of curvature model.
We also apply solar system tests to the models in agreement with 
the cosmological data. 
We find that the exponential of the inverse of the curvature model 
satisfies all the observational tests considered and we derive the allowed 
range of parameters. Current data still allows for small deviations of Einstein gravity. Future, high precision growth data, in combination with expansion history data, will be able to distinguish tiny modifications of standard gravity from the $\Lambda$CDM model.}
\preprint{IFIC/09-66\\
  FTUV-09-1212}

\begin{document} 

\section{Introduction}

 Astronomical observations have led to the inference that our
 universe is approximately  flat and its mass-energy budget
 consists of $5\%$ ordinary matter, $22\%$ non-baryonic 
dark matter, plus a dominant negative-pressure component that 
accelerates the Hubble
expansion~\cite{Dunkley:2008ie,Komatsu:2008hk,Kowalski:2008ez,Tegmark:2006az,Percival:2006gt}. 
The current accelerated expansion of the universe reveals 
new physics missing from our universe's picture, and it 
constitutes the fundamental key to understand the fate of the universe. 

The most economical description of the cosmological parameters 
attributes the negative-pressure \emph{dark energy} component
 to a cosmological constant (CC) in Einstein's equations. 
The CC represents an invariable vacuum energy density that 
assumes a greater importance as the Universe expands.
The equation of state $w$ of the dark energy component 
in the CC case is constant and $w = P_{de}/\rho_{de} = -1$, 
where $P_{de}$ and $\rho_{de}$ denote dark energy pressure 
and energy density, respectively.
However, when computing the vacuum energy density from the quantum 
field theory approach, the naively expected value exceeds the measured one by 123 orders of magnitude 
and it needs to be cancelled by extreme fine-tuning. This is the
so-called CC problem. A related problem is the so called \emph{why now?} 
or \emph{coincidence} problem, i.e. why the dark matter and dark
energy contributions to the energy budget of the universe are similar 
at this precise moment of the cosmic history.

A dynamical alternative attributes the accelerated 
expansion to a cosmic scalar field,
\emph{quintessence}~\cite{Caldwell:1998je,Zlatev:1998tr,Wang:1999fa,Wetterich:1994bg,Peebles:1987ek,Ratra:1987rm}, 
that changes with time 
and varies across space, slowly approaching its ground state.
In this case, the equation of state $w$ could vary over time. 
However, quintessence models are not better than the CC scenario as
regards fine-tuning, since there is no symmetry that explains the 
tiny value of the potential at its ground state.

There exists another possible scenario, in which the gravitational sector 
is modified, as an alternative to explain the observed cosmic acceleration. 
Although this requires the modification of Einstein's equations of
gravity on very large distances~\cite{Dvali:2000hr}, or on small curvatures~\cite{Carroll:2003wy,Capozziello:2003tk,Vollick:2003aw},
this is not unexpected for an effective 
4-dimensional description of higher dimensional
theories. \emph{Modifications of gravity} have been examined 
in the context of accelerated expansion. The proposed modified gravity
models have extra spatial dimensions or an action which is non
linear in the curvature scalar, that is, these models include
extensions of the Einstein-Hilbert action, for instance, 
to higher derivative theories~\cite{Carroll:2004de}, 
scalar-tensor theories or generalized functions of the Ricci scalar $f(R)$. 

Among a plethora of $f(R)$ models, a recent
study~\cite{Amendola:2006we} has identified those
cosmologically acceptable, i.e. models with
a standard matter era followed by an accelerated attractor.
We focus here on the cosmological bounds on these \emph{viable} $f(R)$
models. We use recent SNIa, BAO, CMB and $H(z)$ galaxy ages data
to constrain the background evolution in this class of $f(R)$ models. 
We exploit as well current measurements  
of the linear growth of structure, which provides us an additional 
test to be combined with the background probes. 
The $f(R)$ models which are not ruled out  
by the global cosmological analysis will be examined under solar system 
constraints.

The structure of the paper is as follows. We start in Sec.~\ref{sec:seci} 
specifying the class of modified gravity models explored here, as well as the 
equations which describe the background evolution and the linear perturbation
 theory in a generic $f(R)$ model. Section \ref{sec:data} contains a 
description 
of the different cosmological data sets used in the analysis performed here.
Our results are presented in Sec.~\ref{sec:ana}. 
We describe the solar system constraints in Sec.~\ref{sec:solar}.
We summarize our results, draw our conclusions and discuss future work 
in Sec.~\ref{sec:disc}.

\section{$f(R)$ Models}
\label{sec:seci}
We investigate the simplest family of modified 
gravity models, obtained by adding to the usual Hilbert-Einstein
Lagrangian some function $f(R)$ of the Ricci scalar $R$, 
 with an action given by
\beq
\label{action}
{\cal L} = \int d^4 x \ (R + f(R))  \sqrt{g} + {\cal L}_{matter}~.
\eeq
Here we analyse $f(R)$ models which are cosmologically viable, 
i.e., models which predict a matter dominated period
followed later by an accelerated expansion epoch. 
In the matter domination era the effective equation of state is close 
to $\omega_{eff}=0$ and the scale factor $a$  grows with 
time as $a(t) \sim t^{2/3}$. 

The authors of ~\cite{Amendola:2006we} have explored the general conditions 
for the cosmological viability of $f(R)$ models in the context of a  
flat, homogeneous and isotropic background. The 
cosmological behaviour of $f(R)$ models can be characterized by studying the 
$m(r)$ curve on the $(m,r)$ plane~\cite{Amendola:2006we}, where 
\beq
\label{mr}
m=\frac{R f_{RR}}{1+f_R}~;
\qquad \qquad
r=-\frac{R(1+f_R)}{R+f}~,
\eeq
and $f_R \equiv df/dR$. A given $f(R)$ model will have a standard matter dominated period followed by a late time accelerated era if the conditions $m(r) \approx +0$ and $dm/dr > -1$ at $r \approx -1$ are satisfied, respectively. 

Reference \cite{Amendola:2006we} shows that all $f(R)$ models with an 
accelerated global attractor belong to four classes, 
two of which can be cosmologically acceptable: models of Class II,
asymptotically equivalent to the
$\Lambda$CDM model ($\omega_{eff}=-1$), and models of Class IV, 
which have a non-phantom final accelerated expansion period 
($\omega_{eff}>-1$). In practice, there are not $f(R)$ models belonging 
to Class IV, unless they are built by hand from a well-behaved $m(r)$ function. 
There are other type of models, as those from Class III,
which have an unstable matter era followed by 
a phantom acceleration ($\omega_{eff} < -7.6$). These Class III 
models are generally ruled out by observations, although a more
careful numerical analysis is needed. \\
We focus here on the Class II models of Ref.~\cite{Amendola:2006we}, studying the following four cases:
\noindent
\bea
\label{eq:model1}
H1: &f(R)&= \alpha R^{n}~,\alpha < 0, 0 < n < 1~;\\
\label{eq:model2}
H2: &f(R)&= R \ [\log(\alpha R)]^{q} - R ~, (q>0)~;\\
\label{eq:model3}
H3: &f(R)&= R \  \exp(q/ R) -R~;\\
\label{eq:model4}
H4: &f(R)&= \alpha R^{2}  - \Lambda ~, (\alpha \Lambda \ll 1)~.
\eea

\subsection{Background evolution}

The field equations, which can be obtained varying the action (\ref{action})
with respect to $g_{\mu\nu}$, read

\beq
(1+f_R) R_{\mu\nu}-\frac{g_{\mu\nu}}{2}(R+f-2\Box f_R)
- \nabla_\mu \nabla_\nu f_R = 8 \pi G T_{\mu\nu}~.
\eeq

\begin{figure}[!ht]
\begin{tabular}{c c} 
  \includegraphics[width=7cm]{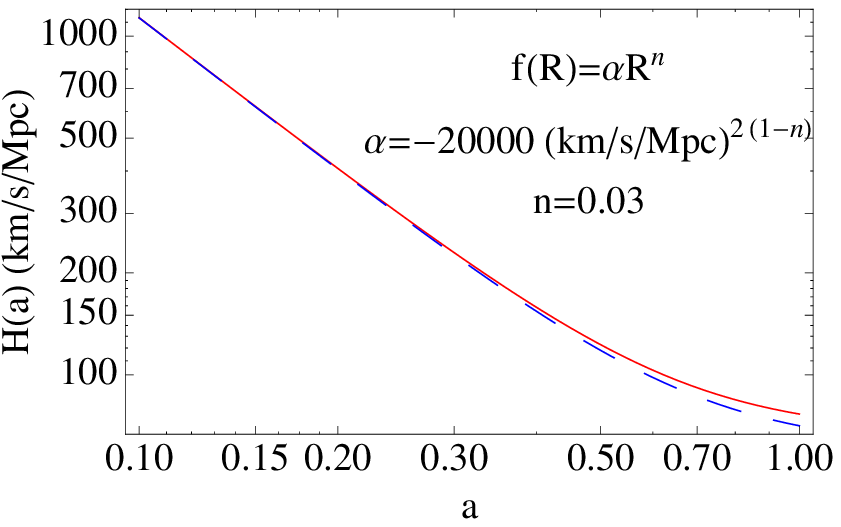}&
 \includegraphics[width=7cm]{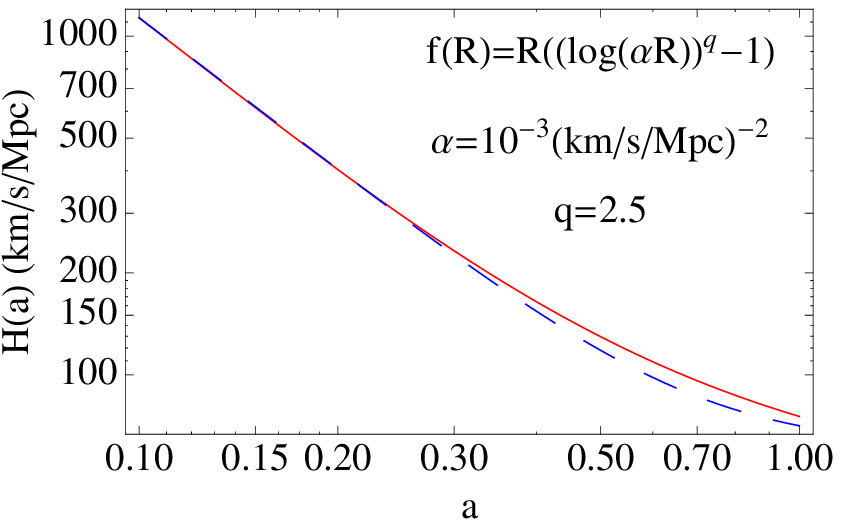}\\
  \includegraphics[width=7cm]{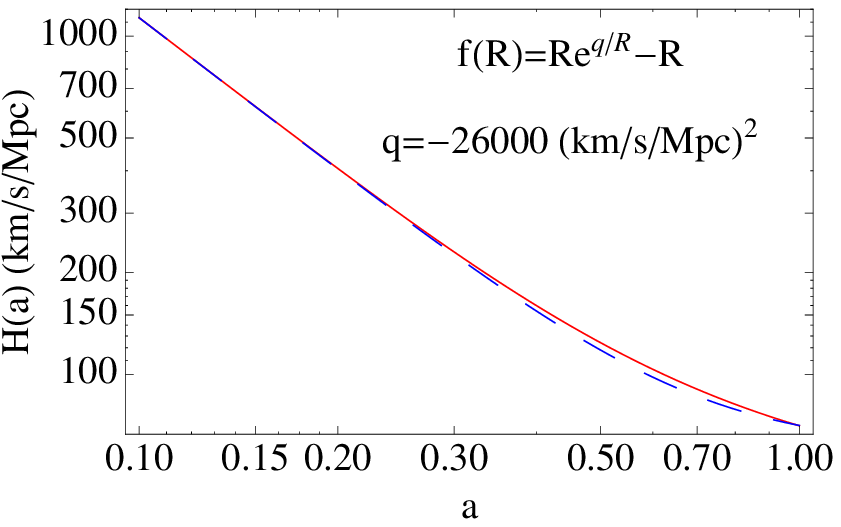}&
 \includegraphics[width=7cm]{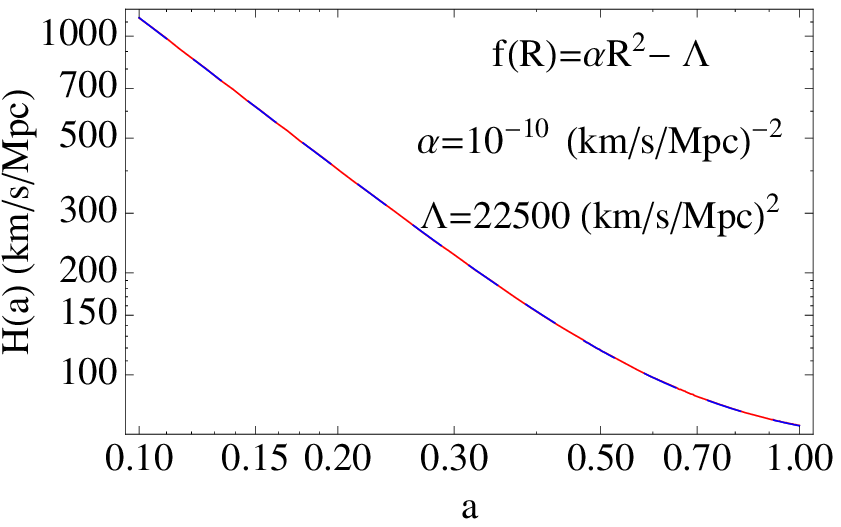}\\
 \end{tabular}
 \caption{{\bf Expansion history of various $f(R)$ models}.
 The red solid curves depict the Hubble rate versus the scale
 factor for the $f(R)$ models explored in this work. 
The parameters were chosen so to have an acceptable expansion history. 
We have added the $\Lambda$CDM model $H(a)$ (blue 
dashed curve) for comparison. }
  \label{fig:hubble}
  \end{figure}

\noindent
The metric we take is of the form of a flat Friedman Robertson Walker (FRW)
background $ds^2 = -dt^2 + a(t)^2\sum_{i=1}^3 (dx^i)^2$, with $a(t)$ the
scale factor.  The Friedmann equation is given by
\beq
\label{backg}
H^2-(H^2+a H H') f_R + a H^2 f_R' + \frac 1 6 f = 
\frac{8\pi G}{3} \rho
\eeq
where $' \equiv d/da$, $H=(da/dt)/a$ denotes the Hubble expansion rate
, $\rho$ refers to the total cold dark matter energy density 
and $R=6(2 H^2+ a H H')$. The present dark matter energy density has 
been fixed to $\Omega_m^0 = 0.24$ (when not treated as a free parameter), accordingly to a recent fit to cosmological 
data~\cite{Komatsu:2008hk}.  
We have integrated numerically the
background equation (\ref{backg}) for the four $f(R)$ Class II 
functions given by Eqs.~(\ref{eq:model1}), (\ref{eq:model2}), (\ref{eq:model3}) and (\ref{eq:model4}). We have determined the ranges of 
the free parameters which lead to a value of the Hubble constant 
within its current 
1$\sigma$ range $H_0 = 74.2\pm 3.6$~km/s/Mpc~\cite{Riess:2009pu}. 
\noindent
Figure \ref{fig:hubble} shows the results for the Hubble parameter
 $H(a)$ for the four different $f(R)$ models explored here. 
As a comparison, we depict as well the Hubble rate for a flat universe
 with a CC ($\Lambda$CDM model)\footnote{The Hubble rate for the 
$\Lambda$CDM model, neglecting the
radiation contribution, reads $H(a)=H_0\sqrt{\Omega_m^0a^{-3}+(1-\Omega_m^0)}$.}. The $H_0$ values for the choice of parameters used in
Fig.~\ref{fig:hubble} are $76.7, 75.4, 70.6$ and $71.0$~km/s/Mpc for 
$f(R)=\alpha R^{n}$, $f(R)=R \left(\log(\alpha R)\right)^{q} - R$, $f(R)=R \exp(q/ R) -R$ and $f(R)=\alpha R^{2}-\Lambda$, respectively.

\subsection{Linear growth rate $\delta(a)$}

\begin{figure}[!ht]
\begin{center}
 \begin{tabular}{c c} 
  \includegraphics[width=7cm]{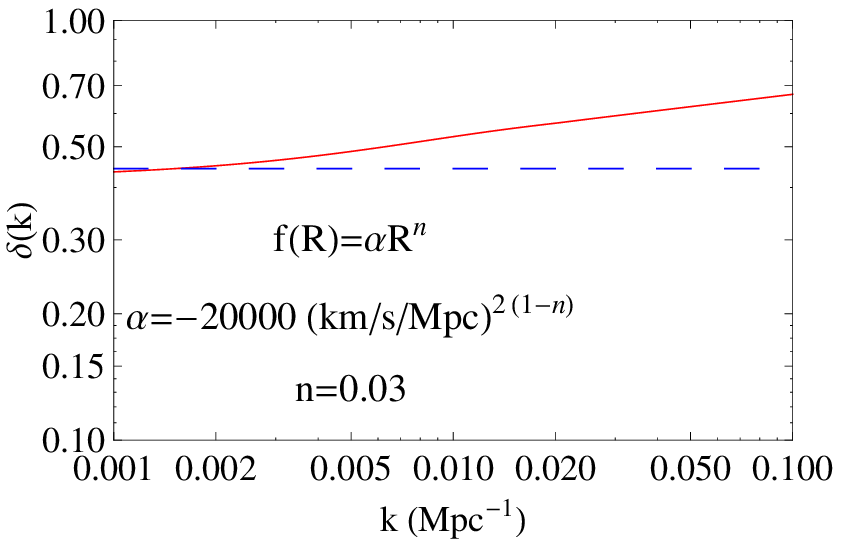}&
  \includegraphics[width=7cm]{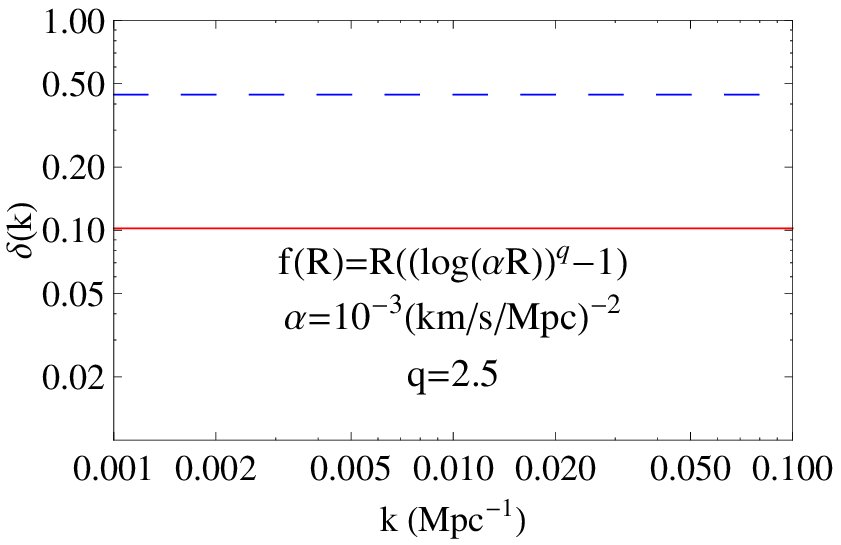}\\
  \includegraphics[width=7cm]{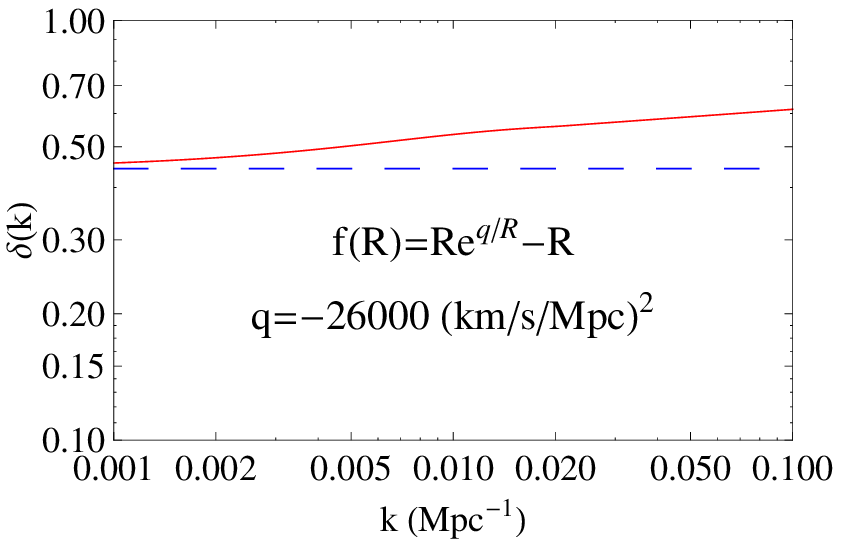}&
  \includegraphics[width=7cm]{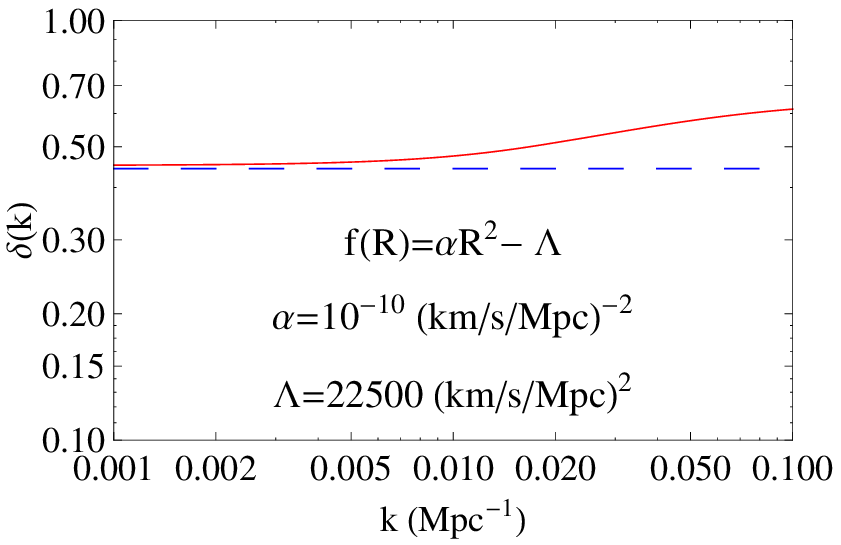}\\
 \end{tabular}
\caption{{\bf Linear growth rate of various $f(R)$ models}.
 The red solid curves depict the present linear overdensity $\delta$ 
as a function of the scale k for the $f(R)$ models explored in this work. 
The parameters were chosen so to have an acceptable 
expansion history. We have added the $\Lambda$CDM model $\delta$(k) (blue dashed curve) 
for comparison.}
  \label{fig:growth}
\end{center} 
\end{figure}

We consider scalar linear perturbations around a flat FRW
background in the Newtonian gauge 

\beq
ds^2 = -(1+2\Psi) dt^2 + a(t)^2(1+2\Phi) \sum_{i=1}^3 (dx^i)^2 \; .
\eeq
The perturbations to the metric are the Newtonian potential $\Psi$ and
the perturbation to the spatial curvature $\Phi$. Since we are working 
in the Jordan frame, in which matter is minimally
coupled, the conservation equations for the cold dark matter component 
have the same form than in general relativity. 
At first order in the perturbations, the conservation equations read 
\begin{eqnarray}
  \label{eq:eomdmw0}
\dot\delta  & = &
3\dot\Phi-\theta~; \nonumber \\
\dot \theta  & = & - H \theta + \left(\frac{k}{a}\right)^2\Psi~,
\end{eqnarray}
where $\dot{}$ means derivative with respect to $t$, $\delta$ is the cold dark matter overdensity and $\theta$ is the dark matter (comoving) peculiar velocity divergence. For subhorizon modes ($k \gsim a H $), and in the quasi-static
limit\footnote{In this limit time derivatives are assumed to be
  negligible with respect to spatial derivatives.}, the perturbed
$0-0$ and $i-j$ ($i\neq j$) components of the Einstein equations read 

\begin{eqnarray}
  \label{eq:einstein2}
2 \left(\frac{k}{a}\right)^2 \left[ \Phi (1+f_{R}) -f_{RR}\left(\frac{k}{a}\right)^2 (\Psi-2
  \Phi)\right]& = & -8 \pi G \rho \ \delta~; \\
\Psi & = & \left(\frac{1-2Q}{1-Q}\right) \Phi~,
\end{eqnarray}
where we have set the anisotropic stress of cold dark matter to zero,
$\rho$ refers to the cold dark matter energy density and we have
neglected the radiation contribution. The factor $Q$ is defined as
\beq
\label{eq:Q}
Q(k,a)= - 2 \; \left(\frac{k}{a}\right)^2 \frac{f_{RR}}{1+f_R}~.
\eeq
\noindent
By substituting the equation for the $i-j$ component into the one for
the $0-0$ component one gets the modified Poisson equation
\beq
\label{modpoisson}
\Phi= \frac{- 8 \pi G}{\left(\frac{k}{a}\right)^2 (1+f_{R})} \rho
\delta \left(\frac{1-Q}{2-3Q}\right)~,
\eeq
which reduces to the standard one if $f_R=0$. The growth factor equation
is obtained by combining Eqs.~(\ref{eq:eomdmw0}) and Eq.~(\ref{modpoisson}), see also Ref.~\cite{Bean:2006up}:
\beq
\label{growth}
\delta'' + \delta' \left( \frac 3 a + \frac{H'}{H} \right)
- \frac{3 \Omega_m(a)}{\left(H/H_0\right)^2 (1+f_R)}
\frac{1-2Q}{2 - 3Q} \frac{\delta}{a^2} = 0~,
\eeq
where $' \equiv d/da$, $\Omega_m(a)=\Omega_m^0 a^{-3}$ and $\delta$ is normalized such that 
$\delta \rightarrow a$ when $a \to 0$. In general relativity, the
factor $Q$ given by Eq.~(\ref{eq:Q}) is zero and therefore the linear
density growth is scale independent for all dark energy
models. However, for $f(R)$ models, the scale dependent $Q(k,a)$ 
induces a nontrivial scale dependence of the growth $\delta$.\\
\noindent
We illustrate this scale dependence of the growth factor in
Fig.~\ref{fig:growth}, where it is shown the present value of the 
matter overdensity $\delta$ as a
function of the scale $k$ for the four $f(R)$ models considered
here. We depict as well the current value of the matter
overdensity for a $\Lambda$CDM universe. Notice that, for the choice of 
parameters which ensure an acceptable $H_0$, 
the growth of matter perturbations within the 
$f(R)=R \ [\log(\alpha R)]^{q}- R$ model is highly suppressed with 
respect to the growth in a universe with a CC. 
For the other three $f(R)$ models the growth is very close to the 
$\Lambda$CDM growth at large scales. However, it shows a $k$ dependence as $k$ increases, due to a larger $Q(k,a)$ factor, see Eq.~(\ref{eq:Q}).
\noindent
Galaxy surveys provide information on $f$, where $f$ is the logarithmic 
derivative of the linear growth rate, i.e. $f\equiv \frac{d ln
  \delta}{d ln a}$. Therefore for our numerical 
analyses we will use $f=\left(\delta'/\delta\right) a$, see the details 
in the next section.

\section{Cosmological data used in the analysis}
\label{sec:data}
In this section we describe the cosmological data used in our numerical analyses\footnote{For practical purposes, in the following, we will use the redshift $z$ instead of the scale factor $a$.}. Four different geometrical probes (SNIa, CMB, BAO and $H(z)$ galaxy ages datasets) will be combined with growth of structure data to derive the cosmological bounds on the free parameters of the $f(R)$ models explored here, see Eqs.~(\ref{eq:model1}), (\ref{eq:model2}), (\ref{eq:model3}) and (\ref{eq:model4}).

\subsection{The Supernova Union Compilation}

The Union Compilation~\cite{Kowalski:2008ez} is a collection of 414 SNIa, which reduces to 307 SNe after selection cuts. It includes the recent large samples of SNIa from the Supernova Legacy Survey and ESSENCE Survey, and the recently extended dataset of distant supernovae observed with the Hubble Space Telescope (HST). In total the Union Compilation presents 307 values of distance moduli ($\mu$), with relative errors, ranging from a redshift $z$ of 0.05 up to $z=1.551$. 
The distance moduli, i.e. the difference between apparent and absolute magnitude of the objects, is given by
\begin{equation}
\mu=5\log\Big(\frac{d_L}{Mpc}\Big)+25~,
\end{equation}
where $d_L(z)$ is the luminosity distance, $d_L(z)=c(1+z)\int_0^z H(z)^{-1}dz$. The $\chi^2$ function used in the analysis reads
\begin{equation}
\chi^2_{SNIa}(c_i)=\sum_z\Big(\frac{(\mu(c_i,z)-\mu_{obs}(z))^2}{\sigma^2_{obs}(z)}\Big) \label{chiw0},
\end{equation}
where, here and in the following, $c_i$ will refer to the free parameters of the f(R) models.

\subsection{CMB first acoustic peak}

We use here the CMB shift parameter $R$, since it is the least model dependent quantity extracted from the CMB power spectrum~\cite{Wang:2006ts}, i.e. it does not depend on the present vale of the Hubble parameter $H_0$. The reduced distance $R$ is written as 
\begin{eqnarray}
R&=&(\Omega_m H_0^2)^{1/2}\int_0^{1089} dz/H(z)~.
\label{eq:rcmb}
\end{eqnarray}  
\noindent
The WMAP-5 year CMB data alone yields $R_0=1.715\pm0.021$ for a fit assuming a 
constant $w$~\cite{Komatsu:2008hk}.
The $\chi^2$ is defined as $\chi^2_{CMB}(c_i) =[(R(c_i)-R_0)/\sigma_{R_0}]^2$. 

\subsection{BAOs}
An independent geometrical probe are BAO measurements. Acoustic oscillations in the photon-baryon plasma are imprinted in the matter distribution. 
These BAOs have been detected in the spatial 
distribution of galaxies by the SDSS~\cite{Eisenstein:2005su} at a redshift 
$z=0.35$ and the 2dF Galaxy Reshift Survey~\cite{Percival:2007yw} at a redshift 
$z=0.2$. The oscillation pattern is characterized by a standard ruler, $s$, 
whose length is the distance sound can travel between the Big Bang and 
recombination and at which the correlation function of dark matter 
(and that of galaxies, clusters) should show a peak. While future BAO data
 is expected to provide independent measurements of the Hubble rate $H(z)$ and of the angular diameter distance $D_A(z)=d_L(z)/(1+z)$ at different redshifts, current BAO 
data does not allow to measure them separately, so they use the spherically correlated function 
\begin{eqnarray} 
D_V(z)&=&\left(D^2_A(z)\frac{c z}{H(z)}\right)^{1/3}~.
\label{eq:bao1}
\end{eqnarray} 
In Ref.~\cite{Sanchez:2007rc}, a tension among SDSS abd BAO datasets was claimed. Therefore, we will focus on the SDSS dataset in the following. The SDSS team reports its BAO measurements in terms of the $A$ parameter,
\begin{eqnarray} 
A(z=0.35)&\equiv& D_V(z=0.35) \frac{\sqrt{\Omega_m H^2_0}}{0.35c}~,
\label{eq:bao}
\end{eqnarray} 
where $A_{SDSS}(z=0.35)=0.469\pm0.017$.
The $\chi^2$ function is defined as $\chi^2_{BAO}(c_i) =[(A(c_i,z=0.35)-A_{SDSS}(z=0.35))/\sigma_{A(z=0.35)}]^2$.

\subsection{Galaxy ages}

We use the $H(z)$ data extracted from galaxy ages in the redshift range $0.1 < z < 1.8$, see Ref.~\cite{Simon:2004tf}. The authors first selected galaxy samples of passively evolving galaxies with
high-quality spectroscopy. Second, they used synthetic stellar
population models to constrain the age of the oldest stars in the
galaxy (after marginalising over the metallicity and star formation
history) and then computed {\em differential} ages and used them as their estimator for $dz/dt$,
which in turn gave $H(z)$. We use the eight data points shown in Figure 1 in 
Ref.~\cite{Simon:2004tf} to test cosmological models by these data sample.
The $\chi^2$ function is defined as
\begin{equation}
\chi^2_{ages}(c_i) =\sum_z \left(\frac{(H(c_i,z)-H(z))^2}{\sigma_{H(z)}^2}\right)~.
\end{equation}

\subsection{Growth factor}
Galaxy surveys measure the redshift of the galaxies, providing, therefore, 
the redshift space galaxy distributions. From those redshifts
the radial position of the galaxy is extracted. However, the inferred
galaxy distribution (and, consequently, the power spectrum) 
is distorted with respect to the true galaxy distribution, because 
in redshift space one neglects the peculiar
velocities of the galaxies. These are the so called \emph{redshift
  space distortions}. 

In linear theory and with a local linear galaxy bias $b$ the relation between the true
spectrum in real space and the spectrum in redshift space reads

\beq
\label{eq:kaiser}
P_\textrm{redshift}(\boldsymbol{k})=
\left(1 + \beta \mu_{\boldsymbol{k}}^2\right)^2 P(\boldsymbol{k}) \ , 
\eeq
where $\beta\equiv f/b$, being $f$ the logarithmic derivative of the growth
factor, and $\mu_{\boldsymbol{k}}$ is the cosine of the angle between the line of 
sight and the
wavevector $\boldsymbol{k}$. Notice that perturbations with
$\boldsymbol{k}$ perpendicular to the line of sight are not distorted.
By averaging over all directions $\mu_{\boldsymbol{k}}$, one obtains the relation
\beq
\label{eq:kaiser2}
P_\textrm{redshift}(k)=\left(1 + \frac{2}{3}\beta +\frac{1}{5} \beta^2
\right) P(k)~.
\eeq
The relation among real space and redshift space overdensities given
by Eq.~(\ref{eq:kaiser})  was first derived by
Kaiser~\cite{Kaiser:1987qv} and it arises from the continuity
equation, which relates the divergence of the peculiar velocity with
the linear matter overdensity.  
Redshift space distortions, then, relate peculiar velocities with
the growth factor $f$. A measurement of $\beta\equiv f/b$ will provide
information on the growth of structure formation if the galaxy bias $b$
is known. One can estimate the redshift distortion parameter $\beta$
both by using the ratio of the redshift space correlation
function to the real space correlation function, see
Eq.~(\ref{eq:kaiser2}) and by exploiting the ratio of the
monopole and quadrupole harmonics of the redshift correlation
function~\cite{1992ApJ...385L...5H}:

\begin{eqnarray}
Q_{\textrm{redshift}}=\frac{P^{(2)}_{\textrm{redshift}}(k)}{P^{(0)}_{\textrm{redshift}}(k)}=\frac{\frac{4}{3} \beta + \frac{4}{7}\beta^2}{1+\frac{2}{3}\beta+\frac{1}{5}\beta^2}~.
\end{eqnarray}
We quote the current available data on $\beta$, the galaxy bias $b$
and the inferred growth factor in Tab.~\ref{tab:tab1}. 
Notice from the first of Eqs.~(\ref{eq:eomdmw0}) that the continuity equation in
$f(R)$ theories is exactly the same than in general
relativity and therefore the relation between peculiar velocities and
the matter overdensity is not modified in the $f(R)$ models studied here.
Consequently, we use the available data on the logarithmic
derivative of the growth factor $f$, see Tab.~\ref{tab:tab1} as an
additional test for $f(R)$ models, to be added to the geometrical
probes previously described.  
The $\chi^2$ function is defined as $\chi^2_{growth} =\sum_j [(f(c_i,z_j,k_0)-f(z_j))/\sigma_{f(z_j)}]^2$. Notice that the theoretical prediction for the growth factor $f(c_i,z_j,k)$ is scale dependent. We choose $k_0=0.1$~Mpc$^{-1}$  to be within the linear regime and within the scale range tested by current surveys, see Tab.~\ref{tab:tab1}.

\begin{table}[bt!]
\begin{center}
\begin{tabular}{c|c|c|c|c} \hline\hline

$z$ & $\beta$ & $b$  & $f$ & References\\
\hline 
0.15&$0.49\pm 0.09$ & $1.04\pm 0.11$&$0.51\pm 0.11$&\cite{Verde:2001sf,Hawkins:2002sg}\\
 \hline
0.35&$0.31\pm 0.04$ &$2.25\pm 0.08$ &$ 0.7\pm 0.18$ &\cite{Tegmark:2006az}\\
 \hline
0.55&$0.45\pm 0.05$ &$1.66\pm 0.35$&$0.75\pm 0.18$&\cite{Ross:2006me}\\
 \hline
0.77&$0.70\pm 0.26$ &$1.30\pm 0.10$ &$0.91\pm 0.36$&\cite{Guzzo:2008ac}
\end{tabular}
\caption{Current available data for the redshift distortion parameter
  $\beta$, the bias $b$ and the inferred growth factor, see Ref.~\cite{Nesseris:2007pa}.}
\label{tab:tab1}
\end{center}
\end{table}

\section{Analysis of cosmological models}
\label{sec:ana}
In this section we present the constraints in four modified gravity models 
(see Eqs.~(\ref{eq:model1}) to (\ref{eq:model4})) which arise from the 
datasets described in the previous section. These models have been shown 
to have a long enough matter domination epoch and late-time accelerated 
expansion~\cite{Amendola:2006we}. 

We show below that the combination of geometrical probes 
(i.e. distance measurements) and growth of structure data 
allows to exclude some models. For those consistent with all the
cosmological data sets used here, we derive the allowed range 
of parameters and discuss the near future improvements. 
We will see in next section that some of the models consistent 
with all cosmological data, are excluded by solar system tests.

In the discussion, we make use of the individual chi-square functions, 
the global chi-square defined by  
$$\chi^2_{tot}(c_i)=\chi^2_{SNIa}(c_i)+\chi^2_{BAO}(c_i)+\chi^2_{CMB}(c_i)+\chi^2_{ages}(c_i)+\chi^2_{growth}(c_i) ,$$ 
and the \emph{only-distances} $\chi^2$, defined as the global chi-square without the last term. If not otherwise specified, the cosmological parameters 
$H_0$ and $\Omega_m$ are fixed to the values $74.2$~Km/s/Mpc and $0.24$ respectively.

\subsection{Model H1:
$f(R) = \alpha R^{n} $ ($\alpha < 0$, $0 < n < 1$)}

This model contains two parameters: the power index of the curvature $n$ 
and the normalization of the modification of gravity $\alpha$.
The $\alpha R^{n}$ model contains the $\Lambda$CDM universe as a limiting case: if $n  \to 0$, then $f(R) \to \alpha$, where the parameter $\alpha$ becomes a cosmological constant. Therefore, it must be allowed by the cosmological data 
within a parameter range. The best fit model is acceptable for all the independent data sets and the full data analysis gives a $\chi^2_{min}$ of 325.3 for 322 d.o.f..

\begin{figure}[!ht]
\begin{center}
\includegraphics[width=8cm]{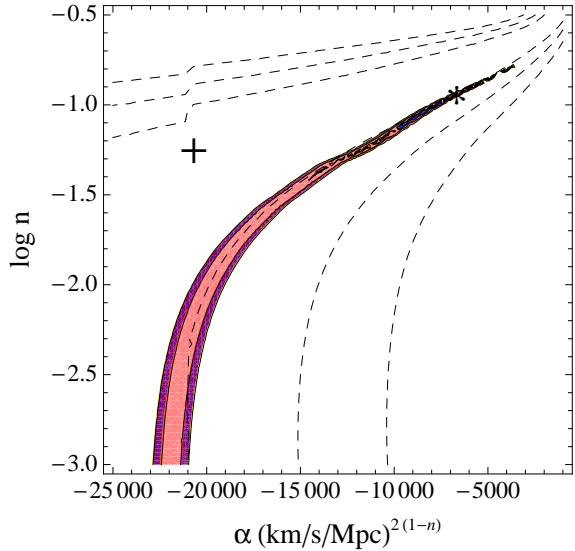}
\caption{{\bf Data analysis of model H1.} Full regions correspond to the 68.3, 95.4 and 99.7 \% CL global analysis allowed regions of parameters n-$\alpha$ of the modified gravity model $f(R) = \alpha R^{n} $. The best fit point of the global analysis is marked with a star. Dashed lines correspond to the 68.3, 95.4 and 99.7 \% CL contours of the growth data analysis. The best fit point of the growth data analysis is marked with a plus sign.}
\label{fig:H1}
\end{center} 
\end{figure}

We firstly discuss the larger $n$ allowed, and how much $\alpha$ deviates from the cosmological constant present in a $\Lambda$CDM universe. 
In Fig.~\ref{fig:H1} we show the $68.3$, $95.4$ and $99.7\%$ CL contours (full colour) resulting from a fit to all the cosmological data exploited here. The global best fit point is marked by a star.
Notice that the power index $n$ can be quite large and that the normalization $\alpha$ can depart from the cosmological constant value. In fact, data prefer $n=0.11$ and $\alpha= -6600$~(km/s/Mpc)$^{2\times(1-0.11)}$, far from the cosmological constant ($\bar{\Lambda}$) limiting case ($\alpha= -2\bar{\Lambda} \sim -20000$~(km/s/Mpc)$^2$ and $n=0$).

The regions allowed by growth of structure data are depicted by dashed lines,  
with the best fit point marked as a plus sign. The allowed regions from a fit  
to distance measurements (i.e. geometrical probes) are depicted by black lines. The statistical power is dominated by distance data: only an expert eye can 
notice the difference among the global analysis allowed regions and those 
coming from the \emph{only-distances} analysis. 
SNIa data are the most important piece of information that constrains 
the modified gravity parameters. 
BAO, CMB and galaxy ages $H(z)$ data sets have a similar weight in 
the statistical analysis. 

Figure \ref{fig:H1} shows some tension between the model predictions and the different data sets for 
 the largest allowed values of $n$. Notice that additional, high precision growth data may further test the high $\alpha$ region and further constrain the deviations of the $f(R) = \alpha R^{n}$ model from a $\Lambda$CDM universe. 

\subsection{Model H2:
$f(R) = R \big(\log(\alpha R)\big)^{q} - R$ ($q>0$)}

This model is described by two parameters: the power index of the logarithm of 
the curvature $q$ and a normalization of the modification of gravity $\alpha$.
The best fit model is acceptable for the all the distance measurements, 
with some tension between the allowed ranges derived from the different 
distance observables. 

\begin{figure}[!ht]
\begin{center}
\includegraphics[width=8cm]{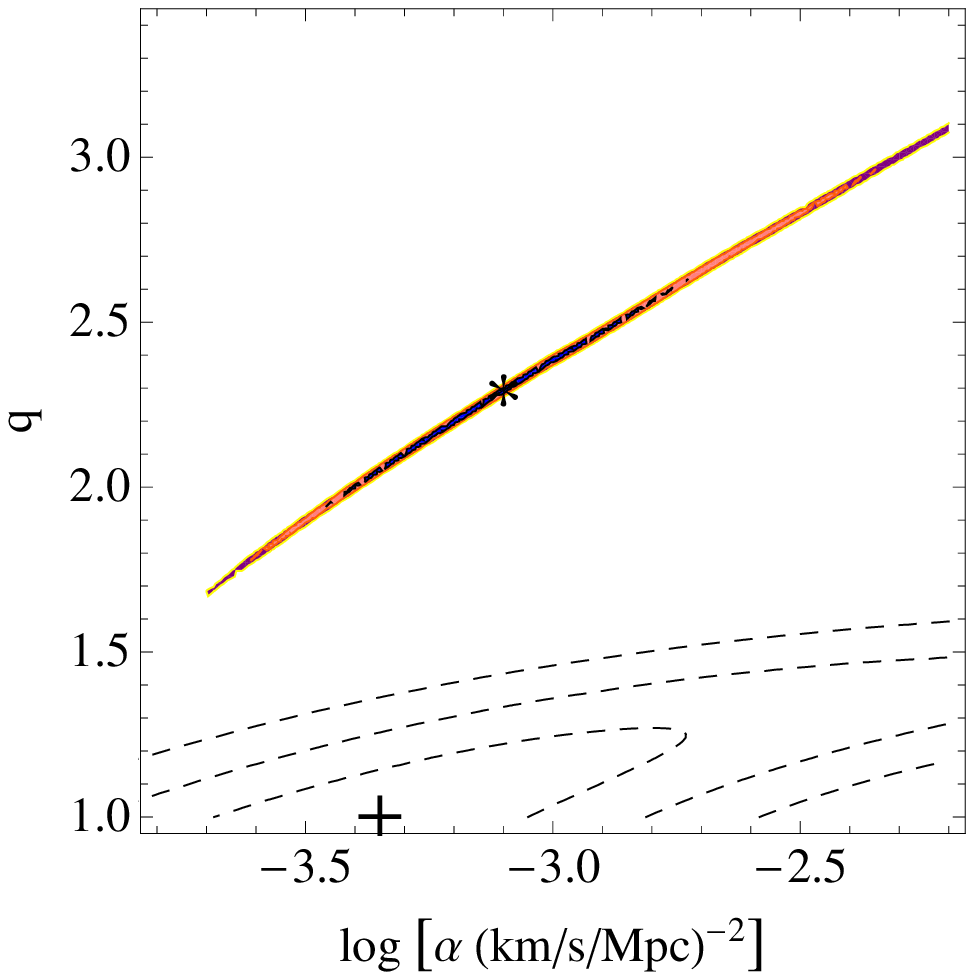}
\caption{{\bf Data analysis of model H2.} Full regions correspond to the 68.3, 95.4 and 99.7 \% CL distances only analysis allowed regions of parameters q-$\alpha$ of the modified gravity model $f(R) = R \big(\log(\alpha R)\big)^{q} - R$. The best fit point of the distances only analysis is marked with a star. Dashed lines correspond to the 68.3, 95.4 and 99.7 \% CL contours of the growth data analysis. The best fit point of the growth data analysis is marked with a plus sign.}
\label{fig:H2b}
\end{center} 
\end{figure}

The relevance of testing this model against cosmological data appears 
when we compare the allowed regions by geometrical probes to those 
coming from a fit to growth of structure data: there is no allowed region 
at more than 99.73\% CL able to fit distances and growth data simultaneously,
see Fig.~\ref{fig:H2b}.

Distance measurements prefer larger values of the power index $q$, 
while growth data prefer a much smaller power index. 
Notice that for parameters that reproduce correctly the expansion history, 
the linear growth is k-independent but very much off the
$\Lambda$CDM model (see Figs.~\ref{fig:hubble} and \ref{fig:growth}). 
Errors in the inferred growth rate are still large 
(see Tab.~\ref{tab:tab1}), but sufficient to test 
this model.
We obtain $\chi^2_{min}=388.7$ for 322 d.o.f., with a probability of the result being due 
to
chance $p< 0.006$. Statistically, we can reject the null hypothesis of the model being 
compatible with data.

\subsection{Model H3: $f(R) = R \exp(q/ R) -R$}

This model contains one free parameter $q$. We also allow here the current 
fraction of the energy density in the form of dark matter, $\Omega^{0}_m$, 
to be an additional free parameter.

The $f(R) = R \exp(q/ R) -R$ model contains the cosmological constant model 
as a limiting case. If $q/R$ is small, then $f(R) \to q$, where the parameter 
$q$ becomes a cosmological constant. The best fit model is acceptable for 
all the independent data sets and the full data analysis gives a 
$\chi^2_{min}$ of 332.4 for 322 d.o.f..

\begin{figure}[!ht]
\begin{center}
\includegraphics[width=8cm]{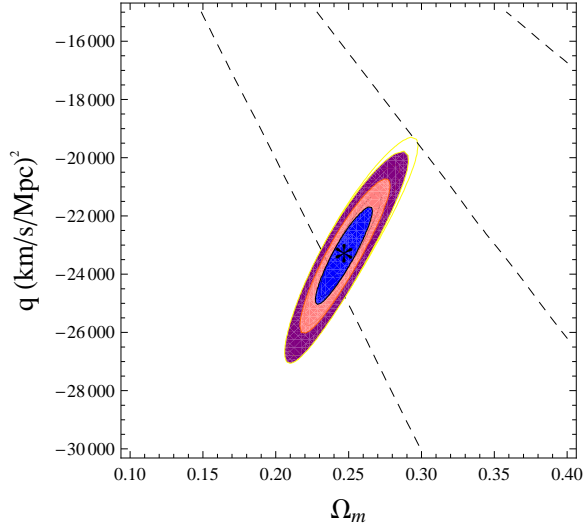}
\caption{{\bf Data analysis of model H3}. Full regions correspond to the 68.3, 95.4 and 99.7 \% CL global analysis allowed regions of parameters q-$\Omega^{0}_m$ of the modified gravity model $f(R) = R \exp(q/ R) -R$. The best fit point of the global analysis is marked with a star. Dashed lines correspond to the 68.3, 95.4 and 99.7 \% CL upper part of the contours of the growth data analysis.}
\label{fig:H3}
\end{center} 
\end{figure}

Figure \ref{fig:H3} depicts the $68.3$, $95.4$ and $99.7\%$ CL contours (full colour) resulting from a fit to all the cosmological data exploited here. The parameter $q$ can deviate from the cosmological constant in $\Lambda$CDM by less than 10\%. 
From the global analysis we obtain $\Omega^{0}_m = 0.245 \pm 0.015$ and $q=-23200 \pm 1200$~(km/s/Mpc)$^2$. The distances only data mostly contributes to strongly constrain the parameter $q$ of this modified gravity model. Notice from Fig.~\ref{fig:H3} that the growth data prefer smaller values of $q$ and $\Omega^{0}_m$, pushing down the global allowed region respect to the distances only allowed region. Not surprisingly, the best fit of the $\chi^2_{growth}$ analysis lies outside the region shown here. Future more accurate growth of structure data could provide tighter bounds on the parameters $q$ and $\Omega^{0}_m$, and potentially reject this model if it would not fit simultaneously distances and future growth data.

\subsection{Model H4: 
$f(R) = \alpha R^{2}  - \Lambda$ ($\alpha \Lambda \ll 1$)}

The $f(R) = \alpha R^{2}  - \Lambda$ ($\alpha \Lambda \ll 1$) model is described by two parameters: the cosmological constant $\Lambda$ and the normalization of the modification of gravity $\alpha$. If $\alpha  \to 0$, then $f(R) \to \Lambda$, a cosmological constant, implying that the model must work in some parameter range. Notice that the Einstein Hilbert action contains the term $R-2 \bar{\Lambda}$, and therefore the $\Lambda$ is twice the usual cosmological constant $\bar{\Lambda}$. 
The best fit model is acceptable for all the independent data sets and the full data analysis gives a $\chi^2_{min}$ of 323.6 for 322 d.o.f.. 

\begin{figure}[!ht]
\begin{center}
\includegraphics[width=8cm]{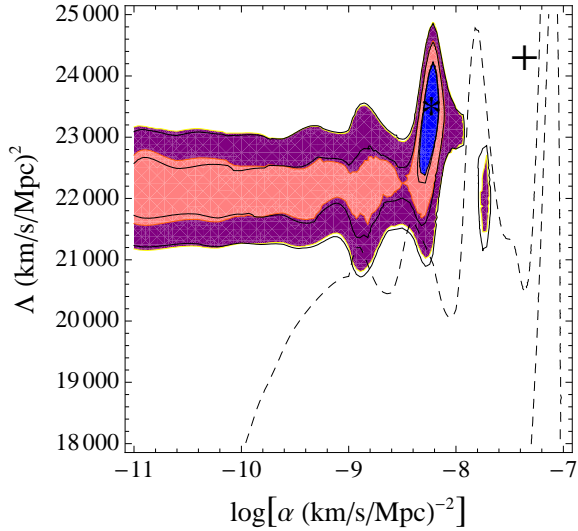}
\caption{{\bf Data analysis of model H4}. Full regions correspond to the 68.3, 95.4 and 99.7 \% CL global analysis allowed regions of parameters n-$\alpha$ of the modified gravity model $f(R) = \alpha R^{2}  - \Lambda (\alpha \Lambda \ll 1)$. The best fit point of the global analysis is marked with a star. Dashed lines correspond to the 68.3, 95.4 and 99.7 \% CL contours of the growth data analysis. The best fit point of the growth data analysis is marked with a plus sign.}
\label{fig:H4}
\end{center} 
\end{figure}

Figure \ref{fig:H4} depicts the $68.3$, $95.4$ and $99.7\%$ CL contours (full colour) resulting from a fit to all the cosmological data exploited here. The data prefer a very small modification of gravity with $\log[\alpha~\textrm{(in km/s/Mpc)}^{-2}] < -8$.
The best fit point of the global analysis is $\log[\alpha~\textrm{(in km/s/Mpc)}^{-2}]= -8.23$ and $\Lambda= 23515$~(km/s/Mpc)$^{2}$, which is about 1-$\sigma$ away from the preferred $\Lambda = 2\bar{\Lambda}$ without a modification of gravity (i.e. $\alpha=0$). High precision future geometrical probes can reduce the model to a negligible perturbation of the $\Lambda$CDM model.

\section{Solar system constraints}
\label{sec:solar}
We explore here the weak field limit of $f(R) = \alpha R^{n}$, $f(R) = \alpha R^{2}  - \Lambda$ and $f(R) = R \exp(q/ R) -R$, modified gravity 
models which have been shown to be consistent with the cosmological probes 
used in the previous section. 

It is well known that $f(R)$ gravity models that produced late time 
acceleration also have problems to pass solar system
tests~\cite{Chiba:2003ir,Navarro:2005gh,Olmo:2005jd,Olmo:2005zr,Capozziello:2005bu,Navarro:2006mw,Hu:2007nk,Olmo:2006eh,Amendola:2007nt,Capozziello:2007ms,Tsujikawa:2008uc}.
The reason is that $f(R)$ gravity theories introduce a scalar 
degree of freedom given by $f_R$ that, for the background cosmological density, is 
very light. As a consequence, it produces a long-range fifth force, 
leading to a dissociation of the space-time curvature from the
local density. Then, the metric around the sun is predicted to 
be different than what is observed. 
Chiba \cite{chiba:2006jp}
has shown under which conditions an $f(R)$ gravity is 
equivalent to a scalar-tensor theory with Parametrized Post-Newtonian 
(PPN) parameter $\gamma=1/2$, far outside the range allowed
by observations, 
$|\gamma -1| < 2.3 10^{-5}$~\cite{Will:2005va}.
  
However, some $f(R)$ theories are still viable: the scalar field mass could be
large and therefore it would not have an effect at solar system scales.
Another possibility is a scale dependent scalar field mass, as in
the chameleon mechanism~\cite{Khoury:2003rn,Cembranos:2005fi,Faulkner:2006ub,Capozziello:2007eu,Brax:2008hh}. In chameleon cosmologies, 
the effective mass of the scalar field becomes very large in 
high density environments (as in the Sun's interior) and the
induced fifth force range would be below the detectability level of
gravitational experiments.
 
In what follows we will apply the criteria presented by Hu and Sawicki
in Ref.~\cite{Hu:2007nk}. If the system's density (for instance, the
Sun's density) changes on scales that are much larger than the scalar field 
Compton wavelength
\begin{equation}
\lambda_{f_{R}}\equiv m^{-1}_{f_{R}}~,
\end{equation}
where
\begin{equation}
m^2_{f_{R}}=\frac{1}{3}\left(\frac{1+f_R}{f_{RR}}-R\right)~,
\end{equation}
the curvature will follow the Sun's density, as in general
relativity. This is the \emph{Compton} condition. 
If it is satisfied at all radii, high densities
will be associated with high curvature and deviations from
general relativity will be highly suppressed. If the \emph{Compton} condition
is not satisfied, one would need then to check the \emph{thin-shell}
condition. This condition applies when the (massive) scalar field is
trapped within the system and its influence is only due to a thin
shell, shielding the fifth force mediated by the scalar field. For the
solar system, the \emph{thin shell} criterion 
reads~\cite{Faulkner:2006ub,Hu:2007nk} 
\begin{equation}
|\Delta f_R(r_\odot)|< (\gamma-1)\frac{G M_\odot}{r_\odot} < 
4.9\times 10^{-11}~,
\end{equation}
where $\Delta f_R$ is the field $f_R$ difference from far inside the
body to very far away. Local gravity constraints for the model H1, 
$f(R) = \alpha R^n$ 
have been already studied in \cite{Amendola:2007nt}, where it is
found that only  $n < 5 \times 10^{-6}$ is allowed. Such values
of $n$  are  too close to $\Lambda$CDM, and therefore not 
interesting cosmologically.

We find that
the only $f(R)$ model which satisfies the \emph{Compton} condition and therefore satisfies
solar system constraints is $f(R) = R \exp(q/ R) -R$. The Compton
wavelength for this model is
$\lambda_{f_{R}}\simeq\sqrt{\frac{3q^2}{R^3}}$. For the best fit
values of $q\sim-20000$~(km/s/Mpc)$^2$ and for densities corresponding
to the solar corona $\rho\simeq10^{-15}$~g/cm$^3$,
$\lambda_{f_{R}}\simeq 10^{-4}r_\odot$. For higher densities, the
Compton wavelength is even smaller and therefore the curvature $R$ will
follow the density profile inside the Sun, as in general relativity. 

The other two $f(R)$ models which survive the cosmological , i.e. $f(R) = \alpha R^{n}$ and $f(R) = \alpha R^{2}  - \Lambda$ do not satisfy the \emph{Compton} nor
the \emph{thin} shell criteria, and therefore they are ruled out by solar 
system observations.

\section{Discussion}
\label{sec:disc}
We have studied a class of modified gravity models, which possess a long enough matter domination epoch and late-time accelerated expansion, as identified by the authors of \cite{Amendola:2006we}. Both the background evolution and the growth of structure have been computed in these well behaved modified gravity cosmologies. We have confronted the expansion history in these cosmologies with SNIa data, the CMB shift parameter $R$, the SDSS BAO measurement and the $H(z)$ data derived from galaxy ages. We have also fitted the linear growth of structure in these $f(R)$ models to the growth information derived from redshift space distortions, as a novel approach. 

Interestingly, we find that the cosmological data exploited 
here possess an enormous potential to rule out modified gravity models. 
$f(R)$ models with a good expansion history like 
$f(R) = R \ \big(\log(\alpha R)\big)^{q} - R$ badly fail to reproduce
the growth measurements and can be statistically rejected with present data.

Other modified gravity models as $f(R) = \alpha R^{n}$, $f(R) = \alpha R^{2}  - \Lambda$ and $f(R) = R \exp(q/ R) -R$ are allowed by all the cosmological data exploited in this study, and include as a limiting case a $\Lambda$CDM universe. 
We show the allowed range of parameters in these models, finding that modifications of gravity must be small. The bounds presented here could be greatly 
improved with future high precision growth data. 

We have also studied the solar system bounds on the three $f(R)$ models which agree with the cosmological data. The  only model which satisfies solar system constraints is $f(R) = R \exp(q/ R) -R$. The other two models, although cosmologically viable, are ruled out by solar system  observations.

From a global fit (which includes SNIa, CMB, BAO, $H(z)$ galaxy ages and growth data) to the $q$ parameter in the exponential model $f(R) = R \exp(q/ R) -R$   and to the current matter energy density  $\Omega^{0}_m$, we obtain $\Omega^{0}_m = 0.245 \pm 0.015$ and $q=-23200 \pm 1200$~(km/s/Mpc)$^2$. Geometrical probes mostly contribute to strongly constrain the parameter $q$ of this modified gravity model, while growth data slightly prefer lower values of both parameters $\Omega^{0}_m$ and $q$. In the parameter allowed region where the exponential model differs from the standard CC cosmology, we find that the growth factor is $k$-dependent, in contrast to the $\Lambda$CDM prediction. More precise growth data at small scales, albeit still within the linear regime, will potentially find small deviations from a universe with a CC.

Our study shows, with the exploration of several $f(R)$ models, that the combination of geometrical probes and growth data offers a powerful tool to rule out modified gravity scenarios and/or constrain deviations from the $\Lambda$CDM picture. We can anticipate that future growth of structure data in the linear regime, combined with a full analysis including the nonlinear regime, will have a very important impact in searching for tiny deviations from Einstein gravity. 

\section*{Acknowledgments}
O.M. and N.R. thank the Fermilab theory group for hospitality.
This work is supported in part by the Spanish MICINN grants 
FPA-2007-60323 and AYA2008-03531, the Consolider Ingenio-2010 project 
CSD2007-00060 and the Generalitat Valenciana grant PROMETEO/2009/116.
The work of O.M. is supported by a Ram\'on y Cajal contract.

\bibliography{bibfr.bib}

\end{document}